\documentclass[aip,amsmath,amssymb,jcp,reprint]{revtex4-1}

\usepackage{graphicx,xcolor}
\usepackage[T1]{fontenc}
\usepackage{siunitx}
\usepackage{extarrows}
\newcommand{\myvec}[1]{\mbox{\boldmath $#1$}}

\definecolor{daidai}{rgb}{1,0.5,0}
\definecolor{usucha}{rgb}{0.75,0.375,0}
\definecolor{kogecha}{rgb}{0.5,0.25,0}
\usepackage[colorlinks=true, linkcolor=usucha, citecolor=daidai, urlcolor=kogecha]{hyperref}

\begin{document}

\title{Theoretical relationship between the macro-texture and micro-structure in dairy processing revealed by the multi-scale simulation of coupled map lattice}

\author{Erika Nozawa}
\email[]{papers@e-rika.net}
\homepage[]{https://www.e-rika.net/english/}
\affiliation{Graduate School of Organic Materials Science, Yamagata University, 4-3-16 Johnan, Yonezawa, Yamagata 992-8510, Japan}

\date{\today}

\begin{abstract}
The theoretical relationship between the macroscopic textural quality and microscopic structural quality appearing in the phase inversion processes from fresh cream via whipped cream to butter is revealed by the multi-scale simulation of coupled map lattice (CML) based on the mesoscopic elementary processes of the emulsion interfaces. Using the Young-Laplace equation, we derive the microscopic particle quantities of the size and density of air bubbles and butter grains in an emulsion from the macroscopic rheological quantities of the overrun and viscosity of the emulsion. In doing so, we focus on the size determined by the ``tug-of-war'' between air bubbles and butter grains via their cohesion pressures, and on the density determined by the ``costume change'' of the emulsion molecular complexes (clad particles, e.g., butter grain-clad air bubbles) to their suitable size. Using the obtained microscopic particle quantities, we now propose a microscopic state diagram, the size-density plane, in addition to the previously proposed macroscopic state diagram, the viscosity-overrun plane. These state diagrams reveal that while the two well-known different phase inversion processes at high and low whipping temperatures appear as the two parallel processes of viscosity dominance and overrun dominance in the viscosity-overrun plane, they appear as the two orthogonal processes of isodensity/size dominance and isosize/density dominance in the size-density plane. This theoretical simulation result is significant for the quality design of butter because it demonstrates that differences in macroscopic textural quality can be easily controlled by differences in microscopic structural quality.
\end{abstract}


\maketitle

\section{Introduction}
\label{introduction}

The fresh cream of an O/W emulsion undergoes phase inversion via whipped cream to the butter of a W/O emulsion by the physical action of whipping. \cite{Fujita,McClements} These multi-component, multi-phase, and multi-scale, complex emulsion systems contain high technical barriers for consistently modeling and simulating the entire phase inversion processes. \cite{Matsumura} In a previous paper, \cite{Nozawa3} we proposed a complex systems model by using the coupled map lattice (CML) \cite{Kaneko3,Kaneko,Yanagitab,Yanagitac,Yanagitad,Nishimori,Nozawa,Nozawa2} to overcome the barriers and succeeded in reproducing entire two different phase inversion processes that appear at high and low whipping temperatures. These two processes each exhibit a series of characteristic spatial patterns (texture patterns), and they were classified by the rheological quantities of the texture patterns into viscosity dominance at high whipping temperatures and overrun dominance at low whipping temperatures on the viscosity-overrun plane. In the butter-formed area on the plane, the butter obtained with viscosity dominance had lower overrun and viscosity, providing a {\it creamy \& soft} texture, while the butter obtained with overrun dominance had higher overrun and viscosity, providing a {\it fluffy \& hard} texture.

The quality design of whipped cream and butter necessitates specifying complementarily textural quality with macroscopic rheological quantities and structural quality with microscopic particle quantities, such as the size of food particles. \cite{Miura,Bourne} Indeed, many studies have been vigorously pursued to measure macroscopic rheological and microscopic particle quantities, and to relate them complementarily. \cite{Hanazawa,Kaneda,Fujii,Kaneda2} To obtain a more direct or theoretical relationship between these quantities, multi-scale simulations that can consistently handle full-scale changes ranging from the micro to the macro level are essential.

This study derives microscopic particle quantities from macroscopic rheological quantities based on mesoscopic elementary processes of emulsion interfaces, which are revealed to be essential for phase inversion by CML multi-scale simulations. We highlight the overrun and viscosity of an emulsion as macroscopic rheological quantities, and the size and density of air bubbles and butter grains in the emulsion as microscopic particle quantities. First, the surface tension of the emulsion, a key factor in quality design, \cite{Matsumura,Hanazawa} is obtained from its overrun and viscosity. Next, the size and density of air bubbles and butter grains are evaluated from the curvature changes at the air bubble-fat globule interface in the emulsion before and after the point when the pressures of them acting on this interface are in balance, using the Young-Laplace equation. \cite{Hanazawa} The size is determined by a ``tug-of-war'' between air bubbles and butter grains via their cohesion pressures at the emulsion interface, and the density is determined by a ``costume change'' of air bubbles and butter grains to their suitable sizes.

Furthermore, this study introduces a new microscopic state diagram called the size-density plane from the obtained size and density of butter grains and discusses the theoretical relationship between the dominance of viscosity and overrun, which specifies the macroscopic textural quality of butter, and the dominance of particle size and density, which specifies its microscopic structural quality. In the viscosity-overrun plane based on macroscopic rheological quantities, viscosity dominance and overrun dominance, which were two parallel processes going from whipped cream to butter, appear in the size-density plane based on microscopic particle quantities as two orthogonal processes: isodensity/size dominance and isosize/density dominance. This is a theoretical result indicating that the structure of the butter obtained is essentially distinct between the two processes. The {\it soft \& creamy} butter obtained by the viscosity dominance at high whipping temperatures is low-density and large-size butter that has undergone the isodensity/size dominance, giving it a {\it soft} texture of low density and a {\it creamy} texture of large size. On the other hand, the {\it hard \& fluffy} butter obtained by the overrun dominance at low whipping temperatures is high-density and small-size butter that has undergone the isosize/density dominance, giving it a {\it hard} texture of high density and a {\it fluffy} texture of small size.

The present paper is organized as follows: Section \ref{model} briefly explains the CML for phase inversion phenomena, as proposed in the previous paper \cite{Nozawa3}. Section \ref{derivation} derives the equations of relationship between the macroscopic rheological quantities (the overrun and viscosity of whipped cream and butter) and the microscopic particle quantities (the size and density of air bubbles and butter grains). Section \ref{changes in microscopic particle quantities} investigates changes in the microscopic particle quantities during the two different phase inversion processes at high and low whipping temperatures by applying the derived equations of relationship to the results of CML multi-scale simulations. Section \ref{state diagrams} discusses a complementary relationship between the macroscopic textural quality and microscopic structural quality of the butter obtained from these two phase inversion processes by introducing a microscopic state diagram, the size-density plane. Section \ref{summary and discussion} contains a summary and discussion.

\section{Model}
\label{model}

We briefly review a coupled map lattice (CML) for simulating phase inversion processes going from fresh cream (O/W emulsion) to butter (W/O emulsion) via whipped cream (foam). \cite{Nozawa3} Hereafter, these colloids, including their intermediate states, are simply referred to as ``emulsions,'' which consist of water, milk fat globules (MFGs; composed of fat droplets, lipid crystals, and membranes), and air. The CML is constructed by a minimal set of procedures that consistently describes multi-combined elementary processes in the phase inversion phenomenon. This set is formulated with appropriately coarse-grained field variables of surface energy, cohesive energy, and velocity of the emulsion, along with parameterized nonlinear maps of whipping, where the emulsion is aerated; coalescence, where the emulsion is changed from partial to complete coalescence; and flocculation, where the emulsion is demulsified.

First, we consider an emulsion filling a relatively flat container and represent it as an emulsion on a two-dimensional square lattice. The lattice points are denoted by $ij$ ($i=0,1,\cdots,N_{x}-1$ and $j=0,1,\cdots,N_{y}-1$), and their position vectors by $\myvec{r}_{ij}=(i,j)=i\myvec{e}_{x}+j\myvec{e}_{y}$. Here, $\myvec{e}_{x}$ and $\myvec{e}_{y}$ are the unit vectors in the $x$- and $y$-axis directions, respectively. An emulsion at lattice point $ij$ is defined as a collection of virtual clad particles that are uniformly distributed in a square cell of size one centered at $ij$ (see the ``particle picture'' in Ref.~\onlinecite{Nozawa}). The clad particles were introduced in the previous paper \cite{Nozawa3} to easily reproduce the complex rheology of water- and air-clad MFGs or MFG-clad water droplets and air bubbles, such as molecular complexes, defined by the framework in which the constituents form a cluster and move together even while they are exhibiting different relaxation behaviors in emulsion flows.

Second, we consider emulsion types appearing in the phase inversion processes by whipping, including fresh cream, whipped cream, and butter, and their constituent MFG types, including fat droplet predominant and lipid crystal predominant, and introduce the surface energy $s_{ij}^{t}$, cohesive energy $c_{ij}^{t}$, and velocity (flow) $\myvec{v}_{ij}^{t}=v_{x\, ij}^{t}\myvec{e}_{x}+v_{y\, ij}^{t}\myvec{e}_{y}$ of the emulsion, and the emulsion energy $h_{ij}^{t}=s_{ij}^{t}+c_{ij}^{t}$ as physical and chemical field variables at lattice point $ij$ at discrete time $t$. Using these field variables, fresh cream is represented as low $s_{ij}^{t}$ and low $c_{ij}^{t}$, whipped cream as high $s_{ij}^{t}$ and low $c_{ij}^{t}$, and butter as low $s_{ij}^{t}$ and high $c_{ij}^{t}$. Also, fat droplet predominant MFGs are represented as low $s_{ij}^{t}$ or low $c_{ij}^{t}$, and lipid crystal predominant MFGs as high $s_{ij}^{t}$ or high $c_{ij}^{t}$. Note that the superscript $t$ denotes the index of time, not the exponent.

Third, we consider only important elementary processes in the physical and chemical changes of the emulsion during the phase inversion, and formulate the whipping procedure $T_{w}$ as a Lagrangian procedure, and the coalescence procedure $T_{c}$ and the flocculation procedure $T_{f}$ as Eulerian procedures. Here, the Lagrangian procedures \cite{Kaneko,Yanagitac} describe the change in field variables in each cell along the flow of virtual (clad) particles under the ``particle picture,'' \cite{Nozawa} while the Eulerian procedures \cite{Kaneko,Yanagitac} describe the change in field variables on each lattice point by local and global interactions under the ``lattice picture.'' \cite{Nozawa}

In the whipping procedure $T_{w}$, whipping-induced flows advect the emulsion viscoelastically while deforming the water-fat globule interface (i.e., MFG membranes). The mesoscale membrane deformation results in a macroscale aeration, which increases the surface energy of the emulsion. Using the clad particle framework, this viscoelastic advection and aeration is formulated as the following maps, which change the surface energy $s_{ij}^{t}$, cohesive energy $c_{ij}^{t}$, and velocity $\myvec{v}_{ij}^{t}$ to $s_{ij}^{*}$, $c_{ij}^{*}$, and $\myvec{v}_{ij}^{*}$, respectively (where the symbol $*$ denotes an intermediate time between discrete times $t$ and $t+1$):
\begin{align}
\label{eq:Tw_s}
s_{ij}^{*}&=\sum_{k=i-1}^{i+1}\sum_{l=j-1}^{j+1}a_{ijkl}^{t}\left\{s_{kl}^{t}+f\left(h_{kl}^{t}\right)g\left(h_{kl}^{t}\right)\Delta s_{kl}^{t}\right\},
\\
\label{eq:Tw_c}
c_{ij}^{*}&=\sum_{k=i-1}^{i+1}\sum_{l=j-1}^{j+1}a_{ijkl}^{t}c_{kl}^{t},
\\
\label{eq:Tw_v}
\myvec{v}_{ij}^{*}&=\sum_{k=i-1}^{i+1}\sum_{l=j-1}^{j+1}a_{ijkl}^{t}\myvec{w}_{kl}^{t},
\end{align}
where $a_{ijkl}^{t}$ is the weight of allocation from lattice point $kl$ to lattice point $ij$, i.e., the fraction of emulsion carried from the cell at $kl$ to the cell at $ij$, and it is given by 
\begin{align}
a_{ijkl}^{t}&=a\left(\myvec{r}_{ij},\tilde{\myvec{r}}_{kl}^{t}\right)=a\left(i,j,\tilde{k}^{t},\tilde{l}^{t}\right)
\nonumber\\
&=\left(\delta_{i\lfloor \tilde{k}^{t}\rfloor}\delta_{j\lfloor \tilde{l}^{t}\rfloor}
+\delta_{i\lfloor \tilde{k}^{t}\rfloor +1}\delta_{j\lfloor \tilde{l}^{t}\rfloor}
+\delta_{i\lfloor \tilde{k}^{t}\rfloor +1}\delta_{j\lfloor \tilde{l}^{t}\rfloor +1}\right.
\nonumber\\
&\left.+\,\delta_{i\lfloor \tilde{k}^{t}\rfloor}\delta_{j\lfloor \tilde{l}^{t}\rfloor +1}\right)
\times\left(1-\left|\tilde{k}^{t}-i\right|\right)\left(1-\left|\tilde{l}^{t}-j\right|\right),
\end{align}
using the Kronecker delta $\delta$ and the floor function $\lfloor\,\rfloor$. \cite{Nozawa} Here, $\tilde{\myvec{r}}_{kl}^{t}=(\tilde{k}^{t},\tilde{l}^{t})$ is the position vector to which the emulsion is advected viscoelastically from the cell at $kl$, and it is given by
\begin{align}
\tilde{\myvec{r}}_{kl}^{t}
&=\tilde{\myvec{r}}\left(\myvec{r}_{kl},\myvec{w}_{kl}^{t},\myvec{v}_{kl}^{t}\right)
\nonumber\\
&=(1-\alpha)\left(\myvec{r}_{kl}+\tau\myvec{w}_{kl}^{t}\right)
+\alpha\left\{\myvec{r}_{kl}+\frac{\tau}{2}\left(\myvec{v}_{kl}^{t}+\myvec{w}_{kl}^{t}\right)\right\}
\nonumber\\
&=\myvec{r}_{kl}+\tau\left\{\frac{\alpha}{2}\myvec{v}_{kl}^{t}+\left(1-\frac{\alpha}{2}\right)\myvec{w}_{kl}^{t}\right\},
\end{align}
with the whipping velocity
\begin{align}
\myvec{w}_{kl}^{t}=\left\{
\begin{aligned}
&\omega\left\{\left(c_{y}-l\right)\myvec{e}_{x}+\left(k-c_{x}\right)\myvec{e}_{y}\right\},&t\bmod\iota=0,\\
&0,&\textrm{otherwise},
\end{aligned}
\right.
\end{align}
around the center of rotation $(c_{x},c_{y})=((N_{x}-1)/2,(N_{y}-1)/2)$. The parameter $\alpha$ is the mixing coefficient representing the mass fraction of MFGs in the emulsion, $\tau$ is the relaxation time of the emulsion velocity $\myvec{v}_{kl}^{t}$, $\omega$ is the angular speed of whipping, and $\iota$ is the time interval of whipping. In addition, $f(h_{kl}^{t})g(h_{kl}^{t})\Delta s_{kl}^{t}$ in Eq.~(\ref{eq:Tw_s}) expresses the surface energy increment due to the macroscale aeration derived from the surface energy increment due to the mesoscale membrane deformation
\begin{align}
\label{eq:ds}
\Delta s_{kl}^{t}
&=\Delta s\left(\myvec{w}_{kl}^{t},\myvec{v}_{kl}^{t}\right)
\nonumber\\
&=\frac{1}{2}\alpha\kappa\left[\left(\myvec{r}_{kl}+\tau\myvec{w}_{kl}^{t}\right)-\left\{\myvec{r}_{kl}+\frac{\tau}{2}\left(\myvec{v}_{kl}^{t}+\myvec{w}_{kl}^{t}\right)\right\}\right]^{2}
\nonumber\\
&=\frac{1}{2}\alpha\kappa\left\{\frac{\tau}{2}\left(\myvec{w}_{kl}^{t}-\myvec{v}_{kl}^{t}\right)\right\}^{2},
\end{align}
using the residual factor as a measure of the ease of aeration
\begin{align}
\label{eq:f(h)}
f(h_{kl}^{t})=\left\{
\begin{aligned}
&\frac{h_{kl}^{t}}{\theta},&h_{kl}^{t}\le\theta,\\
&1,&h_{kl}^{t}>\theta,
\end{aligned}
\right.
\end{align}
and the occurrence factor as a measure of the ease of membrane deformation
\begin{align}
\label{eq:g(h)}
g(h_{kl}^{t})=\frac{1}{1+e^{\beta(h_{kl}^{t}-\theta)}}.
\end{align}
The parameter $\kappa$ is the elastic coefficient of MFG membranes, and $\theta$ and $\beta$ are temperature-dependent threshold and gain coefficient for the surface activity of MFG membranes, respectively.

In the coalescence procedure $T_{c}$, a multiplicative effect between aeration and flocculation of the emulsion causes the partial and then complete coalescence of MFGs, which converts the surface energy of the emulsion to its cohesive energy. This coalescence is formulated as the following maps, which change the surface energy $s_{ij}^{*}$ and cohesive energy $c_{ij}^{*}$ to $s_{ij}^{t+1}$ and $c_{ij}^{t+1}$, respectively:
\begin{align}
\label{eq:Tc_s}
s_{ij}^{t+1}&=\gamma g\left(h_{ij}^{*}\right)s_{ij}^{*},
\\
\label{eq:Tc_c}
c_{ij}^{t+1}&=c_{ij}^{*}+\gamma\left\{1-g\left(h_{ij}^{*}\right)\right\}s_{ij}^{*}.
\end{align}
Here, the occurrence factor $g(h_{ij}^{*})$, which is a measure of the ease of membrane deformation in Eq.~(\ref{eq:g(h)}), becomes a measure of the difficulty of coalescence in Eqs.~(\ref{eq:Tc_s}) and (\ref{eq:Tc_c}). The parameter $\gamma$ is the coalescence coefficient representing the conversion efficiency from surface energy $s_{ij}^{*}$ to cohesive energy $c_{ij}^{*}$.

In the flocculation procedure $T_{f}$, the attractive forces driving flocculation and Ostwald ripening induce the flow of the emulsion in a demulsifying direction, which increases emulsion energy differences at a more macroscopic level that extend between adjacent lattice points. This flocculation is formulated as the following maps, which change the velocity $\myvec{v}_{ij}^{*}$ to $\myvec{v}_{ij}^{t+1}$:
\begin{align}
v_{x\,ij}^{t+1}&=v_{x\,ij}^{*}+\frac{\phi}{2}\left(h_{i+1j}^{t+1}-h_{i-1j}^{t+1}\right),
\\
v_{y\,ij}^{t+1}&=v_{y\,ij}^{*}+\frac{\phi}{2}\left(h_{ij+1}^{t+1}-h_{ij-1}^{t+1}\right),
\end{align}
where the parameter $\phi$ is the flocculation coefficient, and $|\myvec{v}_{ij}^{t+1}|$ is assumed not to exceed one.

Finally, we consider the order of physical and chemical changes of the emulsion in the phase inversion phenomenon, and define the time evolution of surface energy $s_{ij}^{t}$, cohesive energy $c_{ij}^{t}$, and velocity $\myvec{v}_{ij}^{t}$ from discrete time $t$ to $t+1$ as follows:
\begin{align}
&\left\{
\begin{aligned}
&s_{ij}^{t}\rule{0pt}{1.2em} \\
&c_{ij}^{t}\rule{0pt}{1.2em} \\
&\myvec{v}_{ij}^{t}\rule{0pt}{1.2em} \\
\end{aligned}
\right\}
\xlongrightarrow[\substack{\text{Lagrangian}\\ \text{procedure}}]{\substack{\text{\it Whipping}\\ T_{w}}}
\left\{
\begin{aligned}
&s_{ij}^{*}\rule{0pt}{1.2em} \\
&c_{ij}^{*}\rule{0pt}{1.2em} \\
&\myvec{v}_{ij}^{*}\rule{0pt}{1.2em} \\
\end{aligned}
\right\}
\nonumber\\
&\xlongrightarrow[\substack{\text{Eulerian}\\ \text{procedure}}]{\substack{\text{\it Coalescence}\\ T_{c}}}
\left\{
\begin{aligned}
&s_{ij}^{t+1}\rule{0pt}{1.2em} \\
&c_{ij}^{t+1}\rule{0pt}{1.2em} \\
&\myvec{v}_{ij}^{*}\rule{0pt}{1.2em} \\
\end{aligned}
\right\}
\xlongrightarrow[\substack{\text{Eulerian}\\ \text{procedure}}]{\substack{\text{\it Flocculation}\\ T_{f}}}
\left\{
\begin{aligned}
&s_{ij}^{t+1}\rule{0pt}{1.2em} \\
&c_{ij}^{t+1}\rule{0pt}{1.2em} \\
&\myvec{v}_{ij}^{t+1}\rule{0pt}{1.2em} \\
\end{aligned}
\right\}.
\end{align}

In the simulations, we used the following conditions: The lattice size $N_{x}\times N_{y}$ is $100\times 100$, the initial surface energy $s_{ij}^{0}$ is given by a uniform random number within $[0,0.2]$, the initial cohesive energy $c_{ij}^{0}$ is zero, and the initial velocity $\myvec{v}_{ij}^{0}$ is also zero, and the wall boundary conditions are assigned. We also used the following parameter values of nonlinear maps: $\omega=0.01$, $\iota=100$, $\alpha=0.5$, $\tau=1$, $\kappa=2$, $\beta=30$, $\gamma=1$, and $\phi=0.01$. The whipping temperatures were set using temperature-dependent threshold $\theta$ for the surface activity of MFG membranes as follows: $\theta_{H}=0.7$ as a representative temperature (about \SI{17}{\degreeCelsius}) of the high whipping temperature range from 14 to \SI{20}{\degreeCelsius}, and $\theta_{H}=1.4$ as a representative temperature (about \SI{7}{\degreeCelsius}) of the low whipping temperature range from 4 to \SI{10}{\degreeCelsius}. Note that these temperature ranges are those of particular interest in dairy processing because they promote the partial coalescence of the MFGs containing both fat droplets and lipid crystals. \cite{Fujita,Dickinson,Walstra}

\section{Theoretical relationship between the macro-texture and micro-structure of dairy products in phase inversion processes}
\label{theoretical relationship}

\subsection{Derivation of microscopic particle quantities from macroscopic rheological quantities}
\label{derivation}

We will derive microscopic particle quantities from macroscopic rheological quantities. In this study, we will focus on the size and density of air bubbles or butter grains in an emulsion as the former, and on the overrun, i.e., surface energy $s$, and viscosity, i.e., cohesive energy $c$, of the emulsion as the latter. In the following, the derivation of these microscopic particle quantities will be presented in the order of (1) the surface tension $\kappa(h)$ of an emulsion, (2) the size $l_{a}$ of air bubbles and the size $l_{b}$ of butter grains, and (3) the density $d$ of emulsion molecular complexes consisting of air bubbles and butter grains. Since this subsection only discusses the emulsion at a given lattice point at a given time, the superscript $t$ and subscript $ij$ of variables will be omitted for ease of notation (e.g., $x_{ij}^{t}$ as $x$).

\subsubsection{Derivation of surface tension of an emulsion}
\label{derivation of surface tension}

First, we will derive the surface tension of the emulsion, which is a key factor in quality design. \cite{Matsumura,Hanazawa} From Eqs.~(\ref{eq:Tw_s}) and (\ref{eq:ds}), the elastic energy $\Delta u$ ($=f(h)g(h)\Delta s$) of an emulsion interface (the water-air bubble-fat globule interfaces) at emulsion energy $h$ is given by
\begin{align}
\label{eq:du1}
\Delta u
=\Delta u\left(f(h),\kappa,\Delta\myvec{r}\right)
=f(h)\frac{1}{2}\alpha g(h)\kappa\left(\Delta\myvec{r}\right)^{2},
\end{align}
where $\Delta\myvec{r}=(\tau/2)(\myvec{w}-\myvec{v})$ is the displacement of an MFG membrane. Note that the residual factor $f(h)$ represents the fraction of MFG membranes (i.e., springs in parallel with a combined spring constant $\alpha g(h)\kappa$) contributing to aeration. Since the elastic forces of the MFG membrane and emulsion interface are balanced,
\begin{align}
\label{eq:ef=st}
\kappa\Delta\myvec{r}=\kappa(h)\Delta\myvec{r}(h),
\end{align}
where $\kappa(h)$ is the elastic coefficient of the emulsion interface (i.e., the surface tension of the emulsion) and $\Delta\myvec{r}(h)$ is its displacement. By using $\kappa(h)$ and $\Delta\myvec{r}(h)$, and Eq.~(\ref{eq:ef=st}), the elastic energy $\Delta u$ of the emulsion interface is expressed as in
\begin{align}
\label{eq:du2}
\Delta u
&=\Delta u\left(\kappa(h),\Delta\myvec{r}(h)\right)
=\frac{1}{2}\alpha g(h)\kappa(h)\left\{\Delta\myvec{r}(h)\right\}^{2}
\nonumber\\
&=\frac{\kappa}{\kappa(h)}\frac{1}{2}\alpha g(h)\kappa\left(\Delta\myvec{r}\right)^{2}.
\end{align}
Comparing Eqs.~(\ref{eq:du1}) and (\ref{eq:du2}) yields the surface tension of the emulsion,
\begin{align}
\label{eq:k(h)}
\kappa(h)=\frac{\kappa}{f(h)},
\end{align}
as a function of emulsion energy $h$. Substituting Eq.~(\ref{eq:k(h)}) into Eq.~(\ref{eq:ef=st}) also gives the displacement of the emulsion interface,
\begin{align}
\label{eq:dr(h)}
\Delta\myvec{r}(h)=f(h)\Delta\myvec{r}.
\end{align}

\begin{figure}[tbp]
\begin{center}
  \includegraphics[scale=0.5]{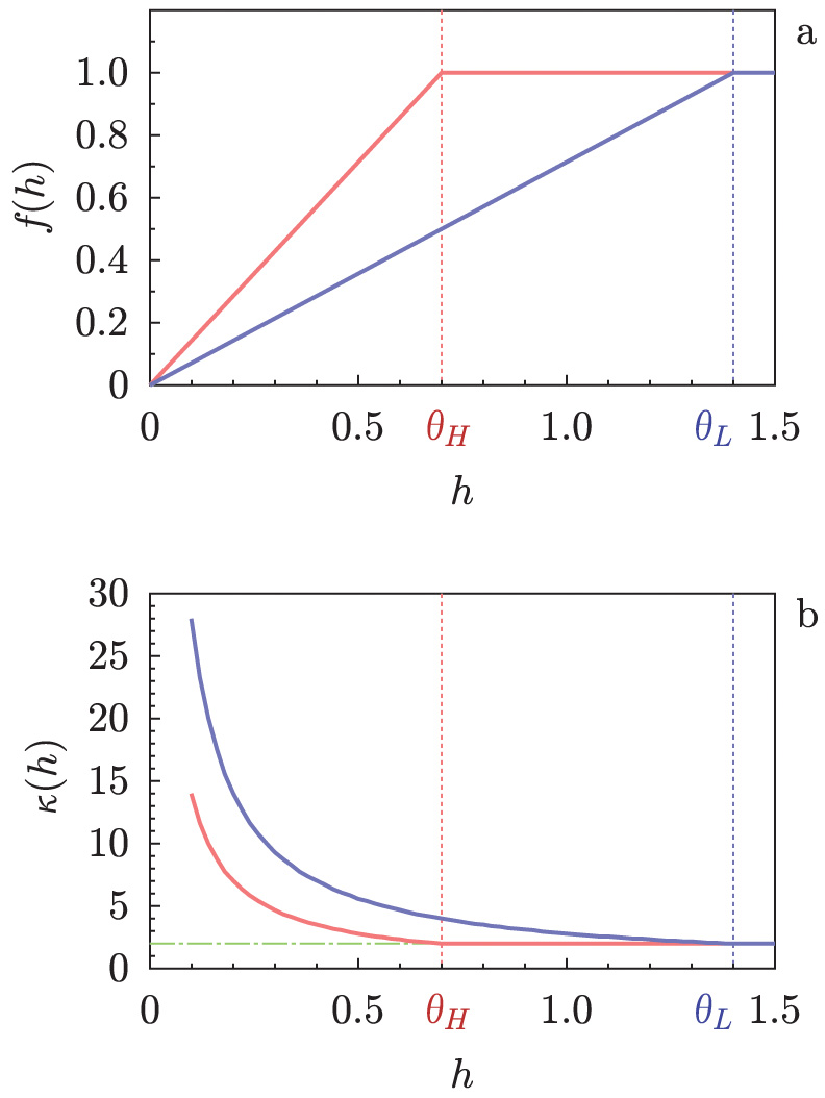}
\end{center}
\caption{(a) Residual fuctor $f(h)$ and (b) surface tension $\kappa(h)$ of the emulsion at emulsion energy $h$. The red lines represent each plot at high whipping temperature and the blue lines at low whipping temperature. The red dotted lines are drawn at $h=\theta_{H}$, the blue dotted lines at $h=\theta_{L}$, and the green dash-dotted line at $\kappa(h)=\kappa$.}
\label{fig:fh_kh.eps}
\end{figure}

Figure~\ref{fig:fh_kh.eps} shows the residual factor $f(h)$ and surface tension $\kappa(h)$ at high whipping temperatures (the red lines, $\theta_{H}=0.7$) and low whipping temperatures (the blue lines, $\theta_{L}=1.4$). In Fig.~\ref{fig:fh_kh.eps}(a), the residual factor $f(h)$ increases proportionally to emulsion energy $h$ and reaches an upper limit of one at the threshold $\theta_{H}$ or $\theta_{L}$ for the surface activity of MFG membranes. On the other hand, in Fig.~\ref{fig:fh_kh.eps}(b), surface tension $\kappa(h)$ decreases inversely proportional to $h$ and reaches a lower limit $\kappa$ (the green dash-dotted line) at $\theta_{H}$ or $\theta_{L}$. These figures show that the increase in the fraction of MFG membranes $f(h)$ contributing to aeration is interpreted as the decrease in surface tension $\kappa(h)$.

As can be seen from Fig.~\ref{fig:fh_kh.eps}(b), the surface tension $\kappa(h)$ and aeration degree $h$ of the emulsion vary depending on the whipping temperature. Here, the aeration degree is represented by emulsion energy $h$ ($=s+c$), since cohesive energy $c$ is small when $h\lesssim\theta$ and then $h$ coincides with surface energy $s$. Under constant aeration degree $h$, surface tension $\kappa(h)$ at low whipping temperatures (the blue curve, ``hard springs'') is higher than that at high whipping temperatures (the red curve, ``soft springs''). Additionally, under constant surface tension $\kappa(h)$, aeration degree $h$ at low whipping temperatures (the blue curve, ``more foaming'') is higher than that at high whipping temperatures (the red curve, ``less foaming''). Therefore, the emulsion interface, which initially has had its hard tension $\kappa(h)$, becomes softer and reduces its tension to $\kappa$ while incorporating a large/small amount of air over a long/short period of time up to aeration degree $h=\theta_{L}$/$=\theta_{H}$ at low/high whipping temperatures.

\subsubsection{Derivation of size of air bubbles and butter grains}
\label{derivation of size}

\begin{figure}[tbp]
\begin{center}
  \includegraphics[scale=0.5]{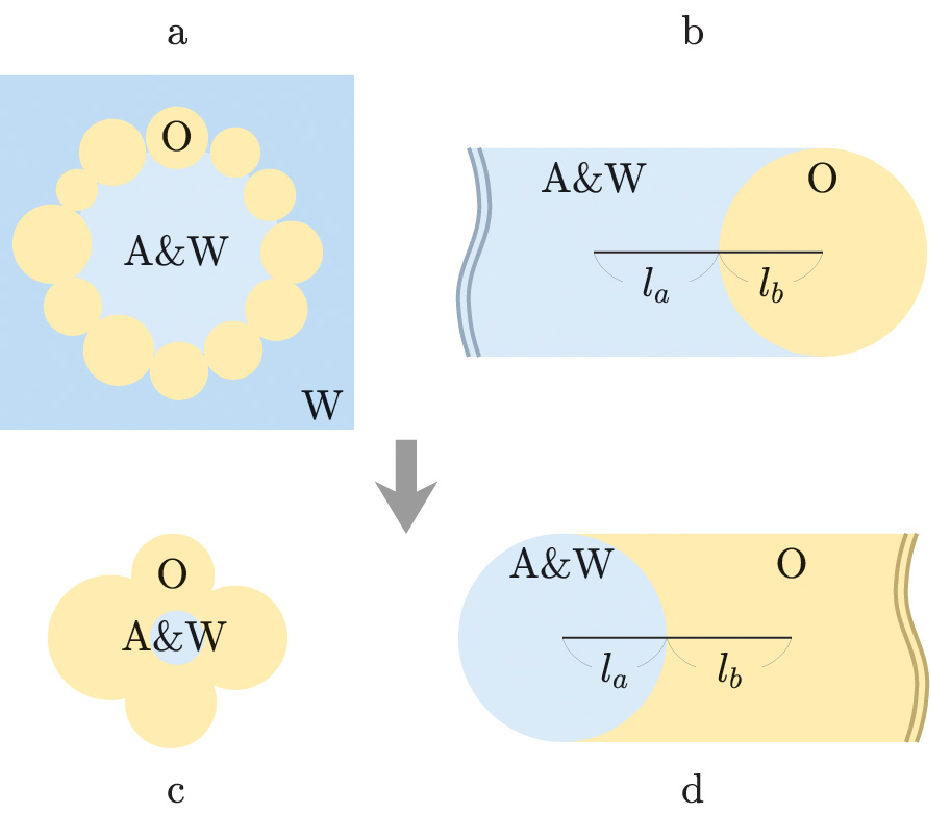}
\end{center}
\caption{Curvature change of the emulsion interface during phase inversion. (a) Schematic illustration of an emulsion molecular complex in whipped cream and (b) its interface. (c) Schematic illustration of an emulsion molecular complex in butter and (b) its interface. A\&W denotes the air and water phase, and O denotes the oil phase. For ease of illustration, the details of the complex were omitted here, although the oil phase (i.e., MFGs) contains fat droplets and lipid crystals, and MFG membranes exist at the interface.}
\label{fig:curvature.eps}
\end{figure}

Next, we will derive the air bubble size $l_{a}$ and the butter grain size $l_{b}$ by using the surface tension $\kappa(h)$ of the emulsion. As shown in Fig.~\ref{fig:curvature.eps}(a), when cream (O/W emulsion) is well whipped, MFGs adsorb and partially coalesce onto air bubbles. \cite{Fujita,Brooker} This results in Fig.~\ref{fig:curvature.eps}(b), where the air and water (A\&W) phase will have negative curvature and the oil (O) phase will have positive curvature. As shown in Fig.~\ref{fig:curvature.eps}(c), when the cream is over whipped, the MFGs flocculate, coalesce, and grow, causing phase inversion to butter (W/O emulsion). \cite{Fujita,McClements} This results in Fig.~\ref{fig:curvature.eps}(d), where the oil (O) phase will have negative curvature and the air and water (A\&W) phase will have positive curvature.

This curvature change of the emulsion interface is evaluated using the Young-Laplace equation as follows:
\begin{align}
\label{eq:YL1}
p_{b}-p_{a}=\kappa(h)\left(\frac{1}{l_{b}}-\frac{1}{l_{a}}\right),
\end{align}
where $p_{a}$ and $p_{b}$ are the pressures of the A\&W and O phases acting on the interface, respectively. The Laplace pressure of the emulsion molecular complex consisting of the A\&W and O phases is also given by
\begin{align}
\label{eq:YL2}
p_{a}+p_{b}=\kappa(h)\left(\frac{1}{l_{a}}+\frac{1}{l_{b}}\right).
\end{align}
From Eqs.~(\ref{eq:YL1}) and (\ref{eq:YL2}), the air bubble size $l_{a}$ and butter grain size $l_{b}$ become
\begin{align}
\label{eq:ll}
l_{a}=\frac{\kappa(h)}{p_{a}},\quad l_{b}=\frac{\kappa(h)}{p_{b}}.
\end{align}

We now relate the microscopic particle quantities, air bubble size $l_{a}$ and butter grain size $l_{b}$, and the macroscopic rheological quantities, overrun $s$ and viscosity $c$, using the pressure acting at the interface as a mediating variable. For whipped cream before curvature change (Fig.~\ref{fig:curvature.eps}(b)), $p_{b}-p_{a}\ge 0$ from $l_{a}\ge l_{b}$, and for butter after curvature change (Fig.~\ref{fig:curvature.eps}(d)), $p_{b}-p_{a}<0$ from $l_{a}<l_{b}$. By using surface energy $s$ and cohesive energy $c$, this sign change is expressed straightforwardly as
\begin{align}
\label{eq:c1}
p_{b}-p_{a}=p_{s}-p_{c},
\end{align}
where $p_{s}=s/1^{3}$ and $p_{c}=c/1^{3}$ with a cell volume of $1^{3}$. The A\&W and O phases of the emulsion molecular complex are formed using emulsion energy $h$, which is simply expressed as
\begin{align}
\label{eq:c2}
p_{a}+p_{b}=p_{h}=p_{s}+p_{c},
\end{align}
where $p_{h}=h/1^{3}$. From Eqs.~(\ref{eq:c1}) and (\ref{eq:c2}), we obtain
\begin{align}
\label{eq:c3}
p_{a}=p_{c},\quad p_{b}=p_{s}.
\end{align}
Substituting Eq.~(\ref{eq:c3}) into Eq.~(\ref{eq:ll}) gives us an important relationship in application between the microscopic particle quantities and the macroscopic rheological quantities as follows:
\begin{align}
\label{eq:lalb}
l_{a}=\frac{\kappa(h)}{p_{c}},\quad l_{b}=\frac{\kappa(h)}{p_{s}}.
\end{align}
Equation (\ref{eq:lalb}) shows that, before and after the point when the two pressures $p_{s}$ and $p_{c}$ are in balance at the emulsion interface, the air bubble size $l_{a}$ is inversely proportional to the cohesion pressure $p_{c}$ of butter grains derived from cohesive energy $c$, and the butter grain size $l_{b}$ to the cohesion pressure $p_{s}$ of air bubbles derived from surface energy $s$. In other words, air bubble size $l_{a}$ and butter grain size $l_{b}$ are determined by a ``tug-of-war'' between air bubbles and butter grains via their respective cohesion pressures $p_{s}$ and $p_{c}$.

\subsubsection{Derivation of density of emulsion molecular complexes}
\label{derivation of density}

Finally, we will derive the number density $d$ of the emulsion molecular complexes consisting of air bubbles and butter grains by using air bubble size $l_{a}$ and butter grain size $l_{b}$. Surface energy $s$ and cohesive energy $c$ are given by
\begin{align}
\label{eq:s}
s&=\left(d\times 1^{3}\right)\kappa(h)\Delta r(h)l_{a},\\
\label{eq:c}
c&=\left(d\times 1^{3}\right)\kappa(h)\Delta r(h)l_{b},
\end{align}
where $d\times 1^{3}$ is the total number of the emulsion molecular complexes in a cell, $\kappa(h)\Delta r(h)$ ($=\kappa(h)|\Delta \myvec{r}(h)|$) is the surface tension force acting on each emulsion molecular complex, and $\kappa(h)\Delta r(h)l_{a}$ and $\kappa(h)\Delta r(h)l_{b}$ are the energies required to form air bubbles of size $l_{a}$ and butter grains of size $l_{b}$, respectively, against the surface tension force. Combining Eqs.~(\ref{eq:lalb}), (\ref{eq:s}), and (\ref{eq:c}) gives us another important relationship between the microscopic particle quantities and the macroscopic rheological quantities as follows:
\begin{align}
\label{eq:d}
d=\frac{p_{s}}{\kappa(h)\Delta r(h)l_{a}}=\frac{p_{c}}{\kappa(h)\Delta r(h)l_{b}}=\frac{1}{\Delta r(h)l_{a}l_{b}}.
\end{align}
Equation (\ref{eq:d}) suggests that the volume of a single emulsion molecular complex is given by the product $\Delta r(h)l_{a}l_{b}$ of three characteristic lengths, including the controllable length $\Delta r(h)$ in dairy processing. In other words, $d$ emulsion molecular complexes (i.e., clad particles) undergo a suitable ``costume change'' following the ratio of three edges $\Delta r(h)$, $l_{a}$, and $l_{b}$ when transforming into states such as whipped cream and butter.

\subsection{Changes in microscopic particle quantities during phase inversion processes from the CML multi-scale simulations}
\label{changes in microscopic particle quantities}

Using the obtained relationship (\ref{eq:lalb}) and (\ref{eq:d}), we will investigate the changes in the average air bubble size $l_{a}$, average butter grain size $l_{b}$, and average density $d$ during two different phase inversion processes, especially going from whipped cream to butter, at high and low whipping temperatures. Here, the macroscopic rheological quantities, overrun $S$ and viscosity $C$, are defined by the total surface energy and total cohesive energy per whipping as follows:
\begin{align}
\label{eq:S}
S&=\frac{1}{\iota}\sum_{t'=t}^{\mathstrut t+\iota-1}\sum_{i=0}^{N_{\mathstrut x}-1}\sum_{j=0}^{N_{\mathstrut y}-1}s_{ij}^{t'},\\
\label{eq:C}
C&=\frac{1}{\iota}\sum_{t'=t}^{\mathstrut t+\iota-1}\sum_{i=0}^{N_{\mathstrut x}-1}\sum_{j=0}^{N_{\mathstrut y}-1}c_{ij}^{t'}.
\end{align}
The calculation of Equations (\ref{eq:lalb}) and (\ref{eq:d}) uses the average surface energy $\bar{s}$ ($=S/(N_{x}N_{y})$), average cohesive energy $\bar{c}$ ($=C/(N_{x}N_{y})$), and average emulsion energy $\bar{h}$ ($=\bar{s}+\bar{c}$) per lattice point. Note that lengths are normalized based on the displacement $\Delta\myvec{r}$ of an MFG membrane (assumed to be about one-tenth of the membrane thickness) as $10^{-3}=$ 1 nm. The actual values are thus predicted to be about air bubble size $l_{a}\times 10^{-6}$ m, butter grain size $l_{b}\times 10^{-6}$ m, and density $d\times 10^{21}$ particles/m$^{3}$.

\subsubsection{Changes in the size of air bubbles}
\label{air bubble size}

\begin{figure}[tbp]
\begin{center}
  \includegraphics[scale=0.5]{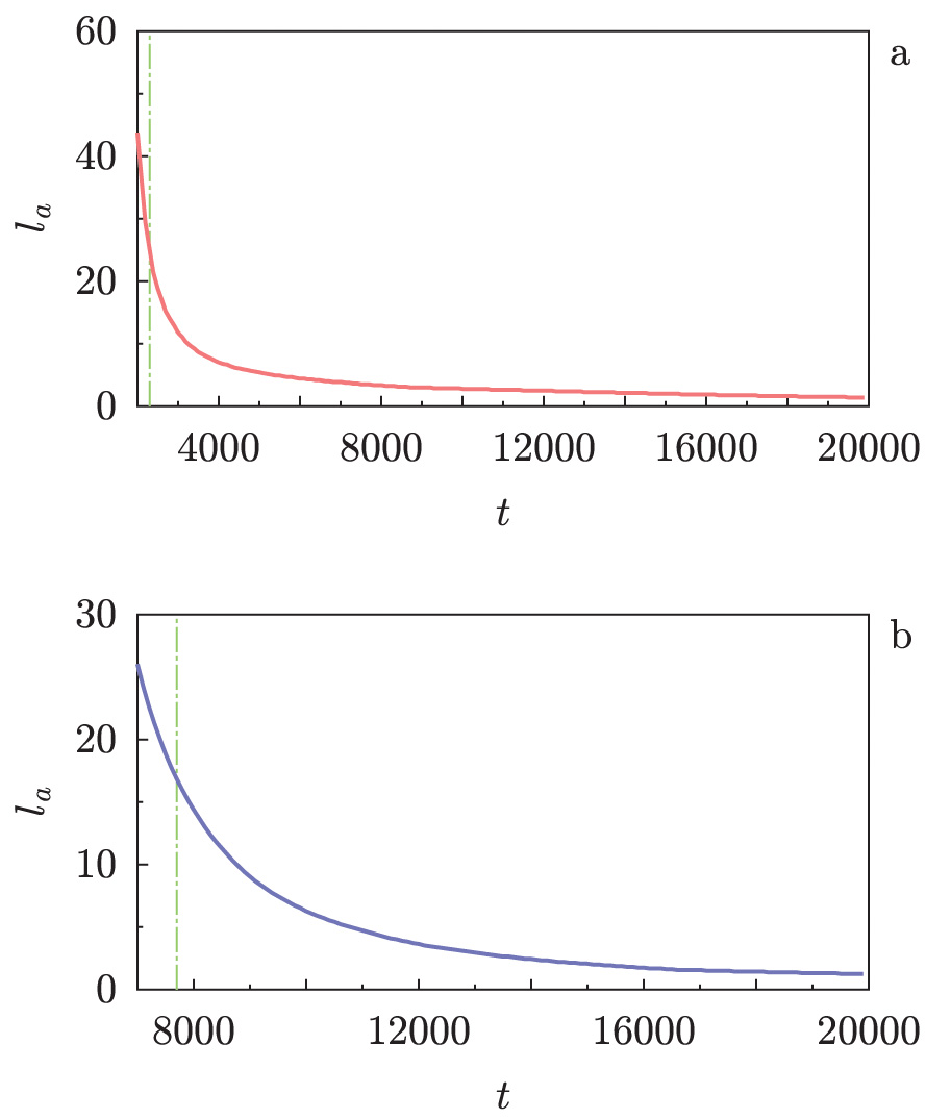}
\end{center}
\caption{(a) Time series of average air bubble size $l_{a}$ at high whipping temperatures, and (b) at low whipping temperatures. The green dash-dotted lines are drawn at the time $t$ when $l_{a}=l_{b}$.}
\label{fig:air_bubble_size.eps}
\end{figure}

\begin{figure}[tbp]
\begin{center}
  \includegraphics[scale=0.5]{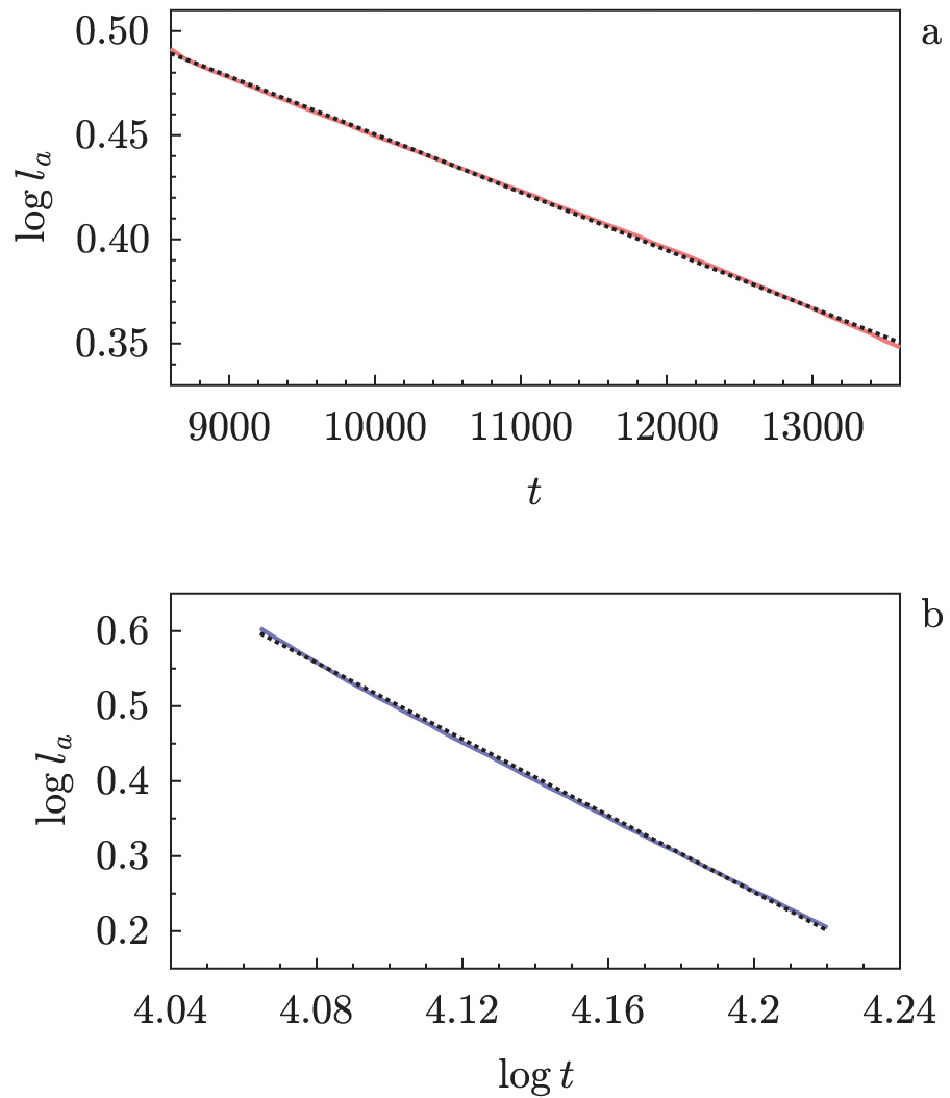}
\end{center}
\caption{(a) Semi-log plot of average air bubble size $l_{a}$ versus $t$ at high whipping temperatures, and (b) log-log plot of $l_{a}$ versus $t$ at low whipping temperatures.}
\label{fig:air_bubble_size_semilog_loglog.eps}
\end{figure}

Figures \ref{fig:air_bubble_size.eps}(a) and \ref{fig:air_bubble_size.eps}(b) show the time series of average air bubble size $l_{a}$ at high and low whipping temperatures, respectively. In both cases, air bubble size $l_{a}$ decreases smoothly, but in distinct ways. As can be seen from the semi-log graph in Fig.~\ref{fig:air_bubble_size_semilog_loglog.eps}(a), air bubble size $l_{a}$ exhibits a single exponential decay at high whipping temperatures. In contrast, as can be seen from the log-log graph in Fig.~\ref{fig:air_bubble_size_semilog_loglog.eps}(b), air bubble size $l_{a}$ exhibits a fractal power-law decay at low whipping temperatures. Note that both graphs were plotted for the time period during which air bubble size $l_{a}$ decreases quasi-stably in the phase inversion processes from whipped cream to butter. These results are in good agreement with the experiments at high and low whipping temperatures. \cite{Jakubczyk,Noda}

The two distinct size decays reveal the following: At high whipping temperatures, air bubbles rapidly decrease in size $l_{a}$ without receiving self-feedback since they are easy to lose the ``tug-of-war'' (Eq.~(\ref{eq:lalb})) with butter grains via cohesion pressure $p_{c}$, while at low whipping temperatures, they slowly decrease in size $l_{a}$ with receiving self-feedback since they are hard to lose the ``tug-of-war.'' This microscopic result, based on the difference in temporal correlations, is also closely related to spatial correlations that generate uniform structures and fractal structures, as will be in Fig.~\ref{fig:butter.eps} in Sec.~\ref{state diagrams}. In addition, this result would provide a useful prediction of the time to reach the end point of whipping \cite{Fujita,Noda}, e.g., around the curvature change point $l_{a}=l_{b}$ (the green dash-dotted lines in Figs.~\ref{fig:air_bubble_size.eps}(a) and \ref{fig:air_bubble_size.eps}(b)), in dairy processing and confectionery.

\begin{figure}[tbp]
\begin{center}
  \includegraphics[scale=0.5]{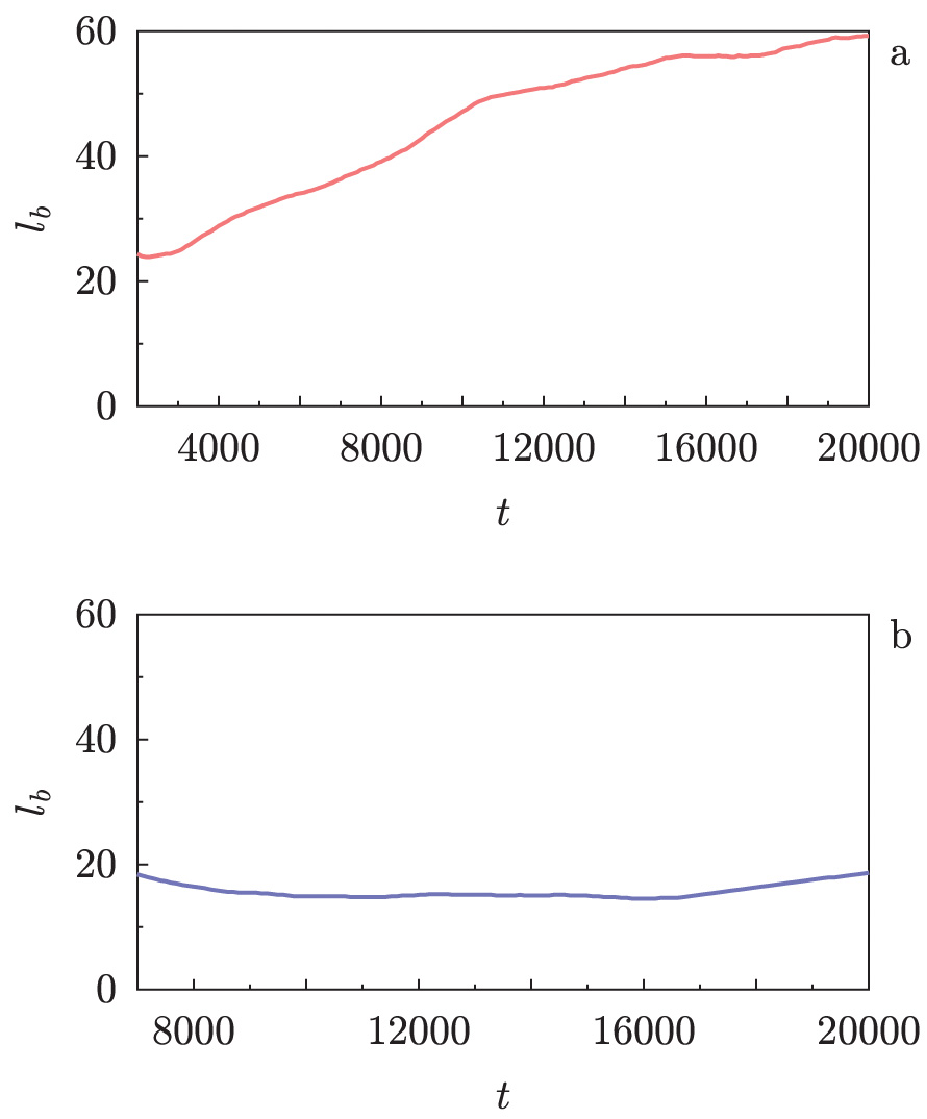}
\end{center}
\caption{(a) Time series of average butter grain size $l_{b}$ at high whipping temperatures, and (b) at low whipping temperatures.}
\label{fig:butter_grain_size.eps}
\end{figure}

\subsubsection{Changes in the size of butter grains}
\label{butter grain size}

Figures \ref{fig:butter_grain_size.eps}(a) and \ref{fig:butter_grain_size.eps}(b) show the time series of average butter grain size $l_{b}$ at high and low whipping temperatures, respectively. As can be seen from both figures, butter grain size $l_{b}$ exhibits very distinct changes depending on whipping temperatures. At high whipping temperatures, butter grain size $l_{b}$ increases by about twice as much. By contrast, at low whipping temperatures, it remains almost constant.

The two distinct size changes reveal the following: At high whipping temperatures, the phase inversion process is size dominance, in which butter grains are formed through coalescence since they easily win the ``tug-of-war'' (Eq.~(\ref{eq:lalb})) with air bubbles via cohesion pressure $p_{s}$, while at low whipping temperatures, it is isosize dominance, in which butter grains are formed without coalescence since they hardly win the ``tug-of-war.'' In addition, the change in butter grain size $l_{b}$ of size dominance exhibits various increase rates, including plateaus, as shown in Fig.~\ref{fig:butter_grain_size.eps}(a), which suggests the growth of butter grains through diverse ways of coalescence.

\subsubsection{Changes in the density of emulsion molecular complexes}
\label{density}

\begin{figure}[!b]
\begin{center}
  \includegraphics[scale=0.5]{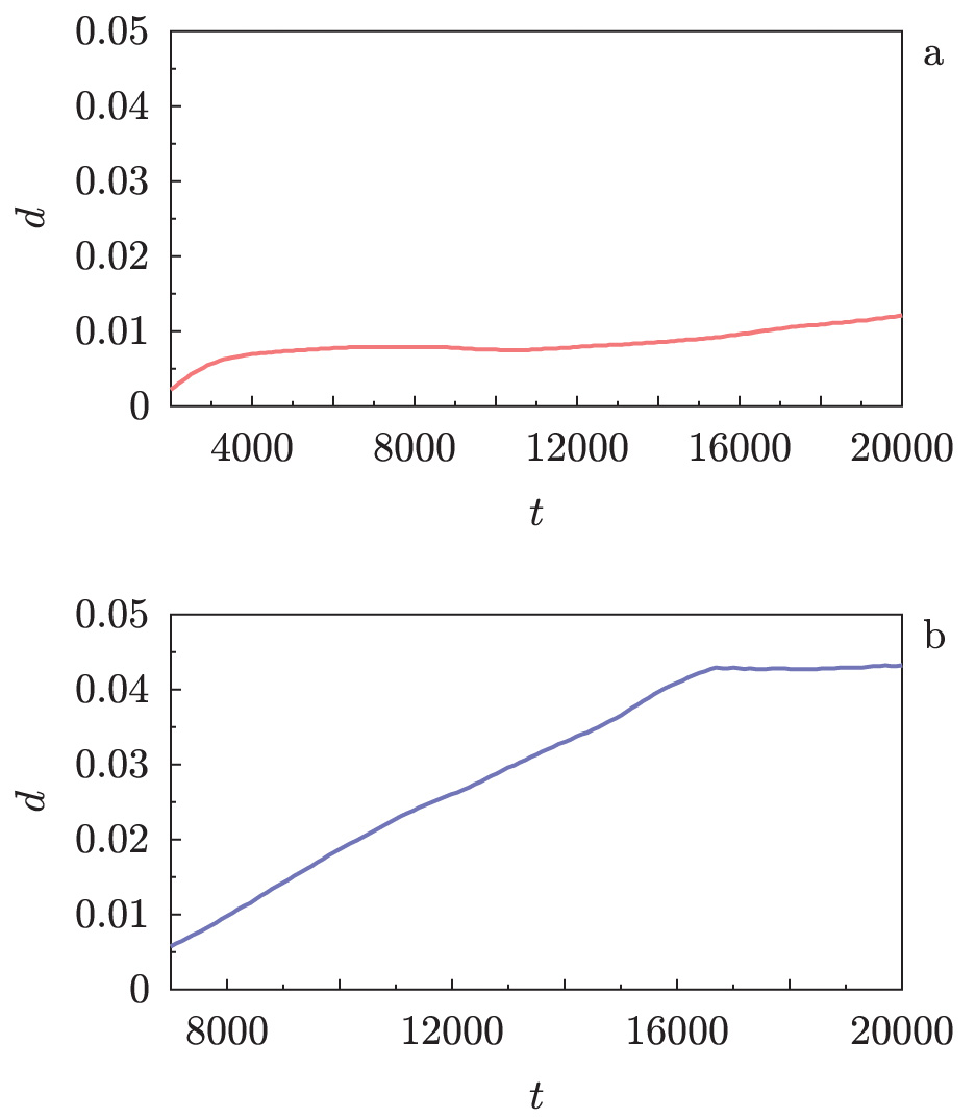}
\end{center}
\caption{(a) Time series of average density $d$ at high whipping temperatures, and (b) at low whipping temperatures.}
\label{fig:molecular_complex_density.eps}
\end{figure}

Figures \ref{fig:molecular_complex_density.eps}(a) and \ref{fig:molecular_complex_density.eps}(b) show the time series of average density $d$ of emulsion molecular complexes consisting of air bubbles and butter grains at high and low whipping temperatures, respectively. As can be seen in both figures, density $d$ also exhibits very distinct changes depending on whipping temperatures. At high whipping temperatures, density $d$ stays almost constant. In contrast, at low whipping temperatures, it rises approximately fourfold.

The two distinct density changes reveal the following: At high whipping temperatures, the phase inversion process is isodensity dominance, in which the emulsion molecular complexes are formed without flocculation since they are constant in volume due to a ``costume change'' (Eq.~(\ref{eq:d})) that is suitable for the decrease in air bubble size $l_{a}$ and increase in butter grain size $l_{b}$. At low whipping temperatures, the phase inversion process is density dominance, in which the emulsion molecular complexes are formed through flocculation since they fall in volume due to a ``costume change'' (Eq.~(\ref{eq:d})) that is suitable for the decrease in air bubble size $l_{a}$ and maintenance of butter grain size $l_{b}$. It is interesting that this result on density $d$ contrasts with that on butter grain size $l_{b}$, which was mentioned in Sec.~\ref{butter grain size}. Note that the changes in air bubble size $l_{a}$ and butter grain size $l_{b}$ are the result of the ``tug-of-war'' over size described in Secs.~\ref{air bubble size} and \ref{butter grain size}.

\subsection{Isodensity/size dominance and isosize/density dominance on a microscopic state diagram}
\label{state diagrams}

\begin{figure}[tbp]
\begin{center}
  \includegraphics[scale=0.5]{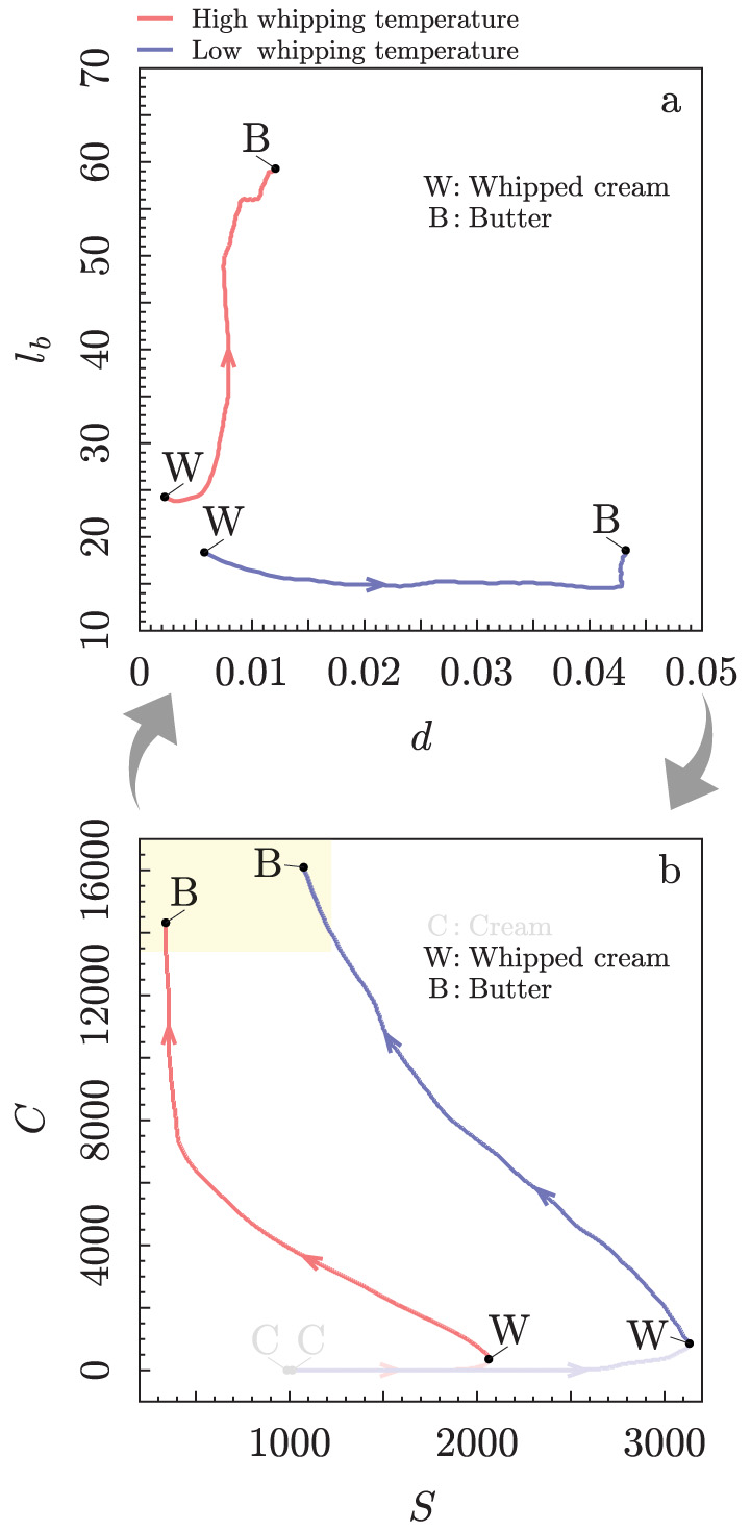}
\end{center}
\caption{Two different phase inversion processes on microscopic and macroscopic state diagrams. (a) Viscosity-overrun ($C$-$S$) plane and (b) size-density ($l_{b}$-$d$) plane. The red and blue curves represent emulsion state changes at high and low whipping temperatures, respectively, which go from start point W (whipped cream) to end point B (butter). The light yellow area represents a butter-formed area.}
\label{fig:cs_ld.eps}
\end{figure}

We propose a microscopic state diagram, the size-density ($l_{b}$-$d$) plane, using average butter grain size $l_{b}$ and average density $d$, in addition to the macroscopic state diagram, viscosity-overrun ($C$-$S$) plane, introduced in the previous paper \cite{Nozawa3}. Figures~\ref{fig:cs_ld.eps}(a) and \ref{fig:cs_ld.eps}(b) show the $l_{b}$-$d$ and $C$-$S$ planes, respectively. In each plane, phase inversion processes are represented as red or blue curves going from start point W (whipped cream) to end point B (butter).

The two different phase inversion processes at high and low whipping temperatures appear as the two {\it parallel} processes in the $C$-$S$ plane based on macroscopic rheological quantities: viscosity dominance (the red curve in Fig.~\ref{fig:cs_ld.eps}(b)) and overrun dominance (the blue curve in Fig.~\ref{fig:cs_ld.eps}(b)). However, in the $l_{b}$-$d$ plane based on microscopic particle quantities, these processes appear as the two {\it orthogonal} processes: isodensity/size dominance (the red curve in Fig.~\ref{fig:cs_ld.eps}(a)) and isosize/density dominance (the blue curve in Fig.~\ref{fig:cs_ld.eps}(a)). This is an important theoretical simulation result for designing butter quality because it demonstrates that underlying the differences in degree of macroscopic textural quality based on rheological quantities (parallelism, or linear dependence) are intrinsic differences in microscopic structural quality based on particle quantities (orthogonality, or linear independence).

\begin{figure}[tbp]
\begin{center}
  \includegraphics[scale=0.5]{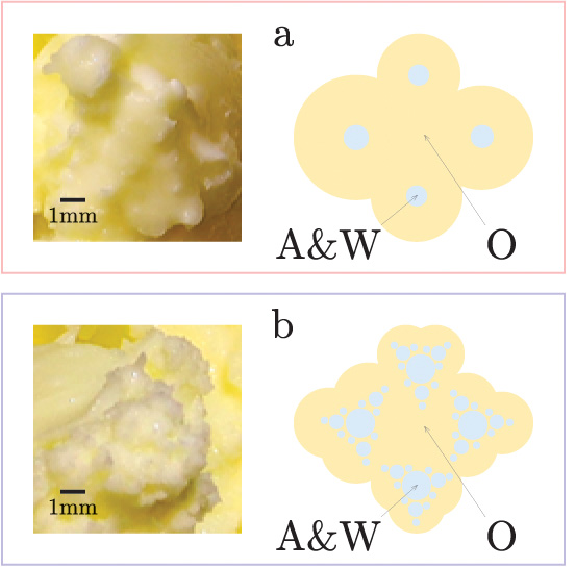}
\end{center}
\caption{(a) Experimental and schematic images of butter grains at high whipping temperatures, and (b) at low whipping temperatures.}
\label{fig:butter.eps}
\end{figure}

We discuss the relationship between the macroscopic textural quality and microscopic structural quality of the butter obtained. First, based on the discussion in Sec.~\ref{changes in microscopic particle quantities}, the microscopic structural quality of butter specified on the $l_{b}$-$d$ plane in Fig.~\ref{fig:cs_ld.eps}(a) is as follows: (1) Butter grains in the isodensity/size dominance at high whipping temperatures have large size $l_{b}$ and low density $d$ (the end point B on the red curve) with low air and water content $l_{a}$, since they are formed via an increase in their size $l_{b}$ through coalescence with exponential decay of air bubble size $l_{a}$, under a constant density $d$ with no flocculation. These butter grains have a uniform structure, as shown in the schematic image in Fig.~\ref{fig:butter.eps}(a). (2) Butter grains in the isosize/density dominance at low whipping temperatures have small size $l_{b}$ and high density $d$ (the end point B on the blue curve) with high air and water content $l_{a}$, since they are formed via an increase in density $d$ through flocculation with power-law decay of air bubble size $l_{a}$, under a constant size $l_{b}$ with no coalescence. These butter grains have a fractal structure, as shown in the schematic image in Fig.~\ref{fig:butter.eps}(b). Note that this microscopic structural quality is also consistent with the empirical fact in dairy processing, because fat droplet predominant MFGs at high whipping temperatures easily cause coalescence (perikinetic mechanism), and lipid crystal predominant MFGs at low whipping temperatures easily cause flocculation (orthokinetic mechanism). \cite{Goff,Smoluchowski}

Next, based on the discussion in the previous paper \cite{Nozawa3}, the macroscopic textural quality of butter specified on the $C$-$S$ plane in Fig.~\ref{fig:cs_ld.eps}(b) is as follows: (1) Butter in the viscosity dominance at high whipping temperatures exhibits lower overrun $S$ and viscosity $C$ (the end point B on the red curve) in the butter-formed area (the light yellow area), i.e., a {\it creamy \& soft} texture, since overrun $S$ decreases rapidly and viscosity $C$ increases gradually in the midst of less foaming. (2) Butter in the overrun dominance at low whipping temperatures exhibits higher overrun $S$ and viscosity $C$ (the end point B on the blue curve) in the butter-formed area (the light yellow area), i.e., a {\it fluffy \& hard} texture, since overrun $S$ decreases slowly and viscosity $C$ increases rapidly in the midst of more foaming.

Finally, the macroscopic textural quality and the microscopic structural quality are related complementarily as follows: (1) The {\it soft \& creamy} butter that is obtained by the viscosity dominance at high whipping temperatures is made from butter grains of low density $d$ and large size $l_{b}$ with low air and water content $l_{a}$ that have undergone the isodensity/size dominance, which give it a {\it soft} texture of low density and a {\it creamy} texture of large size. (2) The {\it hard \& fluffy} butter that is obtained by the overrun dominance at low whipping temperatures is made from butter grains of high density $d$ and small size $l_{b}$ with high air and water content $l_{a}$ that have undergone the isosize/density dominance, which give it a {\it hard} texture of high density and a {\it fluffy} texture of small size.

\section{Summary and discussion}
\label{summary and discussion}

In this paper, we have discussed a theoretical relationship between the macro-texture and micro-structure of whipped cream and butter appearing in the two different phase inversion processes at high and low whipping temperatures. The relationship between the macro-texture and micro-structure was shown in the following three steps. 

First, the microscopic particle quantities of the size and density of air bubbles and butter grains were derived from the macroscopic rheological quantities of the overrun and viscosity of whipped cream and butter. In the derivation, we focused on a ``tug-of-war'' over size between air bubbles and butter grains via the Laplace pressure at the emulsion interface.

Next, distinct differences were demonstrated between changes in these microscopic particle quantities during the two different phase inversion processes at high and low whipping temperatures. Air bubble size exhibits the exponential and power-law decays, butter grain size exhibits the size and isosize dominances, and emulsion molecular complex density exhibits the isodensity and density dominances.

Finally, the macroscopic textural quality specified on the viscosity-overrun plane and microscopic structural quality specified on the size-density plane were related complementarily for the quality of butter obtained from the two different phase inversion processes. The {\it soft \& creamy} butter obtained at high whipping temperatures is made from butter grains of low density and large size, containing a relatively small amount of air and water, while the {\it hard \& fluffy} butter obtained at low whipping temperatures is made from butter grains of high density and small size, containing a relatively large amount of air and water.

As has been discussed in the present paper, the design of whipped cream and butter with a specified quality means how to successfully lead the suitable ``costume change'' of emulsion molecular complexes (clad particles) through various dominances resulting from the ``tug-of-war'' over size. In the future, such designs will be possible using CML-based materials informatics (MI) and process informatics (PI) that combine the macroscopic state diagram ($C$-$S$ plane in Fig.~\ref{fig:cs_ld.eps}(b)) and the microscopic state diagram ($l_{b}$-$d$ plane in Fig.~\ref{fig:cs_ld.eps}(a)).

Indeed, we are currently testing CML-based MI, in which the textural quality requirements (e.g., butter with a {\it fluffy-creamy \& moderately firm} texture) are input on the $C$-$S$ plane using macroscopic rheological quantities (overrun $S^{\star}$ and viscosity $C^{\star}$), and the structural quality specification is output to the $l_{b}$-$d$ plane using microscopic particle quantities (butter grain size $l_{b}^{\star}$ and density $d^{\star}$). This CML-based MI is achieved through a sequence of unit operations provided by CML-based PI. The sequence of unit operations consists of various dominances, such as the viscosity dominance, overrun dominance, isodensity/size dominance, and isosize/density dominance, and is defined as a sequence of selecting dynamical, thermal, and material parameters of CML associated with the whipping force $\kappa\Delta\myvec{r}$ and whipping temperature $\theta$ appearing in $l_{a}$, $l_{b}$, and $d$ in Eqs.~(\ref{eq:lalb}) and (\ref{eq:d}). These CML-based MI and PI will be reported in future papers.

\begin{acknowledgments}
The author would like to thank Professor Tetsuo Deguchi for his valuable comments and insightful suggestions. The author acknowledges the continued support of The Japanese Dairy Science Association Foundation. The author dedicates this work to the memory of Professor Morimasa Tanimoto, who kindly placed his hopes in the role of complex systems science in dairy processing. This research would not have been possible without his support, guidance, and, above all, warm encouragement.
\end{acknowledgments}

\section*{References}
\bibliography{your-bib-file}

@PREAMBLE{
 "\providecommand{\noopsort}[1]{}" 
 # "\providecommand{\singleletter}[1]{#1}%" 
}

@BOOK{Fujita,
   author       = "S. Fujita",
   year         = "2006",
   title        = "Food technology : food emulsions : principles and practice",
   publisher    = "Saiwaishobo",
   address      = "Tokyo",
   note         = "(in Japanese, citation at \\ \url{https://ci.nii.ac.jp/ncid/BA76183579?l=en})"
}

@BOOK{McClements,
   author       = "D. J. McClements",
   year         = "2004",
   title        = "Food Emulsions: Principles, Practices, and Techniques",
   edition      = "2nd",
   publisher    = "CRC Press",
   address      = "Boca Raton",
}

@INBOOK{Kaneko3,
   author       = "K. Kaneko",
   editor       = "K. Kawasaki and M. Suzuki and A. Onuki",
   title        = "Simulating Physics with Coupled Map Lattices",
   booktitle    = "Formation, Dynamics and Statistics of Patterns",
   year         = "1990",
   publisher    = "World Scientific",
   address      = "Singapore",
   pages        = "1--54",
}

@ARTICLE{Noda,
   author       = "M. Noda and Y. Shiinoki",
   year         = "1986",
   title        = "Microstructure and Rheological Behavior of Whipping Cream",
   journal      = "J. Texture Stud.",
   volume       = "17",
   number       = "2",
   pages        = "189",
}

@ARTICLE{Jakubczyk,
   author       = "E. Jakubczyk and K. Niranjan",
   year         = "2006",
   title        = "Transient development of whipped cream properties",
   journal      = "J. Food Eng.",
   volume       = "77",
   number       = "1",
   pages        = "79",
}

@BOOK{Matsumura,
   author       = "Y. Matsumura and K. Matsumiya and A. Ogawa",
   year         = "2017",
   title        = "Technologies for interfacial control and their applications to foods : link the frontiers of research to the actual spots of development",
   edition      = "popular",
   publisher    = "CMC Publishing",
   address      = "Tokyo",
   note         = "(in Japanese, citation at \\ \url{https://ci.nii.ac.jp/ncid/BB24676752?l=en})"
}

@ARTICLE{Kaneda,
   author       = "I. Kaneda",
   year         = "2021",
   title        = "Edible microgel as a texture modifier",
   journal      = "Food Sci. Technol. Res.",
   volume       = "27",
   number       = "5",
   pages        = "687"
}

@ARTICLE{Kaneda2,
   author       = "I. Kaneda and S. Kaneko and Y. Kawabata",
   year         = "2025",
   title        = "Observation of milk clotting behavior using compact NMR instrument",
   journal      = "Milk Science",
   volume       = "74",
   number       = "1",
   pages        = "2",
   note         = "(in Japanese, citation at \\ \url{https://doi.org/10.11465/milk.74.2})"
}

@ARTICLE{Hanazawa,
   author       = "T. Hanazawa and Y. Sakurai and K. Matsumiya and T. Mutoh and Y. Matsumura",
   year         = "2018",
   title        = "Effects of solid fat content in fat particles on their adsorption at the air-water interface",
   journal      = "Food Hydrocoll.",
   volume       = "83",
   pages        = "317"
}

@ARTICLE{Miura,
   author       = "M. Miura",
   year         = "2014",
   title        = "Fascination of Food Rheology 2nd Lecture: Principles of Food Rheology",
   journal      = "Nihon Reoroji Gakkaishi (the Journal of the Society of Rheology, Japan)",
   volume       = "42",
   number       = "3",
   pages        = "215",
   note         = "(in Japanese, citation at \\ \url{https://doi.org/10.1678/rheology.42.215})"
}

@BOOK{Bourne,
   author       = "M. C. Bourne",
   year         = "2002",
   title        = "Food Texture and Viscosity",
   edition      = "2nd",
   publisher    = "Academic Press",
   address      = "San Diego, California"
}

@BOOK{Kaneko,
   author       = "K. Kaneko and I. Tsuda",
   year         = "2001",
   title        = "Complex Systems: Chaos and Beyond",
   publisher    = "Springer-Verlag",
   address      = "Berlin, Heidelberg, New York"
}

@ARTICLE{Yanagitab,
   author       = "T. Yanagita",
   year         = "1992",
   title        = "Coupled map lattice model for boiling",
   journal      = "Phys. Lett. A",
   volume       = "165",
   pages        = "405"
}

@ARTICLE{Yanagitac,
   author       = "T. Yanagita and K. Kaneko",
   year         = "1993",
   title        = "Coupled map lattice model for convection",
   journal      = "Phys. Lett. A",
   volume       = "l75",
   pages        = "415"
}

@ARTICLE{Yanagitad,
   author       = "T. Yanagita and K. Kaneko",
   year         = "1997",
   title        = "Modeling and characterization of cloud dynamics",
   journal      = "Phys. Rev. Lett.",
   volume       = "78",
   pages        = "4297"
}

@ARTICLE{Nishimori,
   author       = "H. Nishimori and N. Ouchi",
   year         = "1993",
   title        = "Formation of ripple patterns and dunes by wind-blown sand",
   journal      = "Phys. Rev. Lett.",
   volume       = "71",
   pages        = "197"
}

@ARTICLE{Nozawa,
   author       = "E. Nozawa",
   year         = "2020",
   title        = "Coupled map lattice for the spiral pattern formation in astronomical objects",
   journal      = "Physica D",
   volume       = "405",
   pages        = "132377"
}

@ARTICLE{Nozawa2,
   author       = "E. Nozawa",
   year         = "2023",
   title        = "Jammed {Keplerian} gas leads to the formation and disappearance of spiral arms in a coupled map lattice for astronomical objects",
   journal      = "Prog. Theor. Exp. Phys.",
   volume       = "2023",
   number       = "6",
   pages        = "063A02"
}

@article{Nozawa3,
   author       = "E. Nozawa and T. Deguchi",
   year         = "2025",
   title        = "Simulating phase inversion processes by coupled map lattice: Toward the theoretical design of food texture and quality in dairy processing from fresh cream to butter via whipped cream",
   journal      = "J. Chem. Phys.",
   volume       = "162",
   number       = "6",
   pages        = "064902"
}

@ARTICLE{Brooker,
   author       = "B. E. Brooker and M. Anderson and A.T. Andrews",
   year         = "1986",
   title        = "The Development of Structure in Whipped Cream",
   journal      = "Food Struct.",
   volume       = "5",
   number       = "2",
   pages        = "Article 12"
}

@BOOK{Dickinson,
   author       = "E. Dickinson and D. J. McClements",
   year         = "1996",
   title        = "Advances in food colloids",
   publisher    = "Chapman \& Hall",
   address      = "London"
}

@INBOOK{Walstra,
   author       = "P. Walstra and {van Vliet}, T. and W. Kloek",
   editor       = "P.F. Fox",
   title        = "Crystallization and rheological properties of milk fat spreads.",
   booktitle    = "Advanced dairy chemistry (2nd ed.) Vol.2: Lipids",
   year         = "1994",
   publisher    = "Chapman \& Hall",
   address      = "London",
   pages        = "179--211",
}

@ARTICLE{Goff,
   author       = "H. D. Goff",
   year         = "1997",
   title        = "Instability and partial coalescence in whippable dairy emulsions",
   journal      = "J. Dairy. Sci.",
   volume       = "80",
   number       = "10",
   pages        = "2620"
}

@ARTICLE{Smoluchowski,
   author       = "M. v. Smoluchowski",
   year         = "1917",
   title        = "Versuch einer mathematischen Theorie der Koagulationskinetik kolloider L{\"o}sungen",
   journal      = "Zeitschrift f. physik. Chemie",
   volume       = "92",
   number       = "1",
   pages        = "129"
}

@ARTICLE{Fujii,
   author       = "K. Sekiguchi and M. Tanimoto and S. Fujii",
   year         = "2023",
   title        = "Mesoscopic Characterization of the Early Stage of the Glucono-$\delta$-Lactone-Induced Gelation of Milk via Image Analysis Techniques",
   journal      = "Gels",
   volume       = "9",
   number       = "3",
   pages        = "202"
}

\end{document}